# Influence of telopeptides on the structural and physical properties of polymeric and monomeric acid-soluble type I collagen

Róisín Holmes,[a,b] Steve Kirk,[b] Giuseppe Tronci,[a,c] Xeubin Yang [a] and David Wood [a*]

[a] Department of Oral Biology, Faculty of Medicine, University of Leeds, Wellcome Trust Brenner Building, St James' University Hospital, Leeds, LS9, 7TF, United Kingdom

[b] Southern Lights Biomaterials, Marton, 4710, New Zealand

[c] Textile Technology Research Group, School of Design, University of Leeds, Leeds, LS2 9JT, UK

* Corresponding author. Tel.: 0113 3436160. E-mail address: d.j.wood@leeds.ac.uk

## Abstract

Currently two factors hinder the use of collagen as building block of regenerative devices: the potential antigenicity, and limited mechanical strength in aqueous environment. Polymeric collagen is naturally found in the cross-linked state and is mechanically tougher than the monomeric, cross-link-free (acid-soluble) collagen *ex vivo*. The antigenicity of collagen, on the other hand, is mainly ascribed to inter-species variations in amino acid sequences, which are primarily located in the non-helical terminal telopeptides. Although these can be removed through enzymatic treatment to produce atelocollagen, the effect of telopeptide removal on triple helix organization, amino acidic composition and thermal properties is often disregarded. Here, we compare the structural, chemical and physical properties of polymeric and monomeric type I collagen with and without telopeptides, in an effort to elucidate the influence that either covalent crosslinks or telopeptides possess. Circular dichroism (CD) was used to examine the triple helical conformation and quantify the denaturation temperature ($T_d$) of both monomeric collagen (36.5 °C) and monomeric atelocollagen (35.5 °C). CD measurements were combined with differential scanning calorimetry (DSC) in order to gain insight into the triple helix-to-coil thermal transition and shrinkage temperature ($T_s$) of polymeric atelo collagen (44.8 °C), polymeric collagen (62.7 °C), monomeric atelo collagen (51.4 °C) and monomeric collagen (66.5 °C). Structural and thermal analyses were combined with high pressure liquid chromatography (HPLC) to determine the content of specific collagen amino acidic residues used as markers for the presence of telopeptides and mature crosslinks. Hydroxylamine was used as the marker for polymeric collagen, and had a total content of 9.66% for both polymeric and polymeric atelocollagen; tyrosine was used as the marker for telopeptide cleavage, was expressed as 0.526% of the content of polymeric collagen and the partially-reduced content of 0.39%.

## 1. Introduction

The biocompatibility and versatility of collagen derived products for medical applications has long been recognized and reflected in its wide scale research (Cao and Xu, 2008). As a scaffold material, it benefits from good biodegradability, affinity for biomolecules and exhibits organizational and macromolecular properties similar to the natural extracellular matrix (ECM) (Drury and Mooney, 2003). Its wide application includes sutures, tissue replacement and regeneration, cosmetic surgery, dental membranes and skin regeneration templates (Ferreira et al., 2012). Collagen has a unique triple helix structure made of three left-handed polypeptide (α-chains) chains held together by hydrogen bonds between the peptide bond (NH) of a glycine residue with the peptide carbonyl (C=O) of the adjacent polypeptide (Lodish et al., 2000). Each chain follows the amino acid motif, -Gly-X-Y-, where X and Y are often proline and hydroxyproline (Lodish, 2000). Glycine (Gly) is the smallest essential amino acid, and provides the H atom which is able to fit inside the crowded center of the triple helix to form the hydrogen bonds. The fixed angle of the C-N bond is due to the high content of peptidyl-proline or peptidyl-hydroxyproline bonds which allow the rotational freedom needed to form the tight packaging of the three chains, thus organizing the triple helix (Walters and Stegemann, 2014, Lodish et al., 2000). The short segments of the C- and N- termini of the polypeptides (the telopeptides), don't possess the repeating Gly-X-Y motif and are therefore non-helical (Parenteau-Bareil et al., 2010, Hanai and Sawada, 2012). Instead, the end telopeptides possess an uncommon amino acid hydroxylysine which is important for the formation of collagen fibrils (Lodish et al., 2000). The fibrils are stabilized by covalent aldol cross-links between lysine-lysine or lysine- hydroxylysine from the C- terminus to the N- terminus of adjacent molecules, thus stabilizing the side packing (Lodish et al., 2000, Shoulders and Raines, 2009). These fibrils possess high tensile strength with a diameter between 50 – 200 nm, and when packed side-by side in tissues such as tendon, form collagen fibers (Li et al., 2013, Lodish et al., 2000).

The use of biological material for medical applications requires a distinction between immunogenicity and antigenicity. Immunogenicity is about triggering an immune response, whilst antigenicity can be determined by macromolecular features of an antigen molecule such as three-dimensional (3D) conformation and amino acid sequence (Lynn et al., 2004).

Collagen, as an animal-derived biomaterial has always raised concerns regarding its potential to evoke immune responses (Parenteau-Bareil et al., 2010, S, 1995, Furthmayr and Timpl, 1976, R, 1976). However, the interpretation of immunochemical reactions to collagen-containing implants is often complicated by the presence of cell remnants, or chemicals from extraction or cross-linking treatments (DeLustro and Condell, 1986, Allaire and Guettier, 1994, M and Lyman, 1990).

## 1.1. Atelocollagen

The collagen molecule can be divided into three domains: the terminal amino (N-) telopeptide, the triple helix, and the terminal carboxy (C-) telopeptide (Morimoto et al., 2014). Treatment of collagen with proteolytic enzymes (e.g. pepsin or ficin) can cleave the terminal telopeptides to produce atelocollagen with an intact triple helical conformation (Lodish, 2000).

It has been claimed that the majority of collagen antigenicity is attributed to the terminal telopeptides, however, the biological effects of atelocollagen are not yet fully understood (Lynn et al., 2004). The antigen determinants for collagen can be classified in three categories: helical recognition by antibodies dependent on 3D conformation located within the triple helical portion of native collagen, amino acid sequence and terminal located in the non- helical telopeptides of the molecule. It had been eluded that collagen devoid of terminal telopeptides can eliminate its immunogenicity (Lynn et al., 2004, Miyata and Taira, 1992). The removal of telopeptides results in an amorphous arrangement of collagen molecules and a loss of the collagen fibril pattern in the reconstituted product, due to the roles of the C- and N-terminus telopeptides in cross-linking and fibril formation (Lynn et al., 2004, Amani et al., 2014) The induced positively charged surface of the atelo collagen can significantly increase its solubility and therefore processabilty of the collagen as a biomaterial.

## 1.2. Polymeric collagen

All collagenous tissues contain a fraction of soluble monomeric collagen which is extractable in weak acidic solutions such as acetic acid (17.4 mmol). In mature tissues, such as tendons, the bulk of the collagen consists of insoluble, highly cross-linked polymerized fibers of type I collagen (polymeric collagen) with a smaller amount as acid-soluble monomeric collagen (<10 %) (Steven and Torrre-Blanco, 1975, Wong et al., 2014). The natural cross-links are chemically rearranged with age to form acid-stable aldminine cross-links, which provide increased mechanical strength of the tissue (Orgel et al., 2011, Wong et al., 2014). The aim of the present study was to assess the influence of telopeptide removal on the structural and physical properties of polymeric and monomeric type I collagen.

## 2. Methods

### 2.1 Materials

Monomeric atelocollagen (Collagen Solutions, UK), Polymeric collagen (Southern Lights Biomaterials, New Zealand), Polymeric atelocollagen (Southern Lights Biomaterials, New Zealand).

### 2.2 Chemical and structural characterization

Circular dichroism (CD) spectra of collagen samples were acquired (ChirascanCD spectrometer, Applied Photophysics Ltd) using 0.2 mg·ml$^{-1}$ solutions in HCl (10 mM). A homogenizer was used to dissolve polymeric and atelo polymeric collagen. Sample solutions were collected in quartz cells of 1.0 mm path length, whereby CD spectra were obtained with 4.3 nm band width and 20 nm·min$^{-1}$ scanning speed. A spectrum of the HCl (10 mM) solution was subtracted from each sample spectrum.

$$\theta_{mrw,\lambda} = \frac{MRW \times \theta_\lambda}{10 \times d \times c} \quad (1)$$

Where $\theta_\lambda$ is observed molar ellipticity (degrees) at wavelength $\lambda$, $d$ is pathlength (1 cm) and $c$ is the concentration (0.0002 g·ml$^{-1}$) (Kelly and Price, 2000).

A temperature ramp was conducted from 20 to 60 °C with 20 °C·hour$^{-1}$ heating rate with ellipticity measurements at 221 nm fixed wavelength. The 221 nm coincides with the positive band associated with the collagen triple helix and its destruction will be related to a lower value of ellipticity. The denaturation temperature ($T_d$) was determined as the mid-point of thermal transition.

Differential Scanning Calorimetry (DSC) was used in order to investigate the thermal denaturation ($T_m$) of collagen samples (TA Instruments Thermal Analysis 2000 System and 910 Differential Scanning Calorimeter cell base). DSC temperature scans were conducted with 10-200 °C temperature range and 10 °C·min$^{-1}$ heating rate. 5-10 mg sample weight was applied in each measurement and three scans were used for each sample formulation. The DSC cell was calibrated using indium with 20 °C·min$^{-1}$ heating rate under nitrogen atmosphere.

High pressure liquid chromatography (HPLC) was used to investigate the amino acid occurrence in collagen samples (Dionex Ultimate 3000 HPLC, Dionex Softron GmbH, Germany). For acid stable amino acids, hydrolysis was performed in HCl (6 M) for 24 hours at 110 °C in an evacuated sealed tube followed by fluorescence detection. Results were calculated as residues per 1000 residues.

**2.3 Physical characterization**

Scanning electron microscopy (SEM) was used for microscopic analysis of collagen by gold-coating the samples in order to examine the fibrillary meshwork and banding pattern (Tronci et al., 2012). Samples were mounted onto 10 mm stubs and electron micrographs captured (FEI Quanta 600) via backscattered electron detection at 10 kV and 12 – 13 mm working distance.

## 3. Results and discussion

### 3.1. Chemical and structural characterization

CD is defined as the unequal absorption of left-handed and right-handed circularly polarized light. When the chromophores of the amides of the polypeptide backbone of proteins are aligned in arrays, their optical transitions are shifted or split into multiple transitions (Greenfield, 2006). The spectra of proteins are dependent on their conformation, so CD is a useful tool to estimate the structure of unknown proteins and monitor conformational changes due to denaturation (Greenfield, 2006). Far-UV CD spectra of monomeric type I collagen displayed a positive band at 221 nm and a negative band at 198 nm, characteristic of the triple helical conformation which decreases during denaturation (figure 1 a).

The magnitude ratio of the positive to negative band (RPN) in monomeric type I collagen spectra) was found to be 0.094 comparable to the literature value of native collagen, RPN: 0.117 (Tronci et al., 2015).

The monomeric atelocollagen had an RPN value of 0.189. The positive band at 221 nm at similar molar ellipticity to monomeric collagen, represents an intact triple helical conformation, despite the removal of the telopetides and the covalent inter-strand cross-links they provided.

Polymeric type I collagen displayed a wide positive band centered at 214 nm, characteristic of random coils and no negative band. The change in the absorption wavelengths is evidence for alterations in the secondary structure, in terms of electronic transitions in the chain backbone or in the helically arrayed side groups of collagen {Menders, 2002 #203}. The spectra could imply that the natural cross-links found in the mechanically stronger polymeric collagen, hinder the coiled supramolecular assembly characteristic of monomeric collagen.

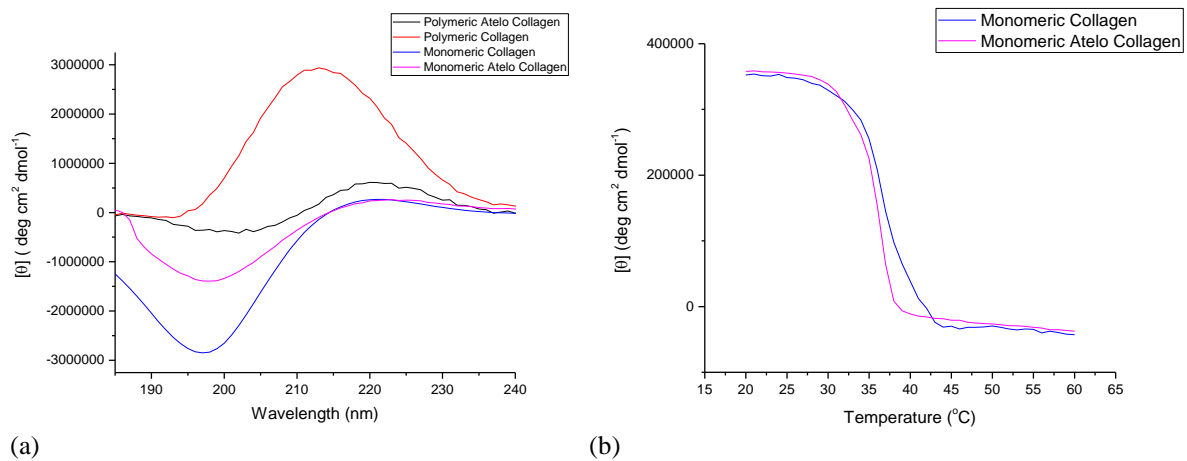

**Fig. 1.** (a) CD graph polymeric atelo, polymeric, monomeric atelo and monomeric type I collagen; (b) temperature ramp graph to show denaturation of monomeric collagen and monomeric atelo collagen.

The spectra of polymeric atelo collagen (RPN: 1.21 (native monomeric collagen (RPN: 0.117) bore no resemblance to the broad positive band displayed for polymeric collagen, and instead showed a positive band (221 nm) similar to monomeric atelocollagen. This could suggest that the enzyme-catalyzed procedure to cleave the telopeptides, also resulted in the destruction of the natural cross-links that differentiate polymeric collagen from monomeric collagen. However, the difference between the molar ellipticity of the positive and negative peak for polymeric atelocollagen suggests some denaturation of the triple helical structure.

A temperature ramp between 20 and 60 °C was used to follow the denaturation of collagen triple helices to randomly coiled form. This could not be performed on polymeric collagen due to the lack of the characteristic 221 nm band maximum. Instead, it was used as a tool to examine the influence of telopeptides on the denaturation temperature of monomeric collagen (figure 1.b).

Denaturation temperature ($T_d$) was measured at half the initial molar ellipticity of the characteristic positive band (221 nm). For monomeric collagen, $T_d$ was 36.5 °C and monomeric atelocollagen, $T_d$ was 35.5 °C.

These differing values could be due to the subdomain of collagen, and the roles that the C- and N- telopeptides provide for intermolecular covalent cross-links which help to stabilize the collagen triple helix; without these additional interactions, the atelocollagen triple helices denature at a lower temperature.

DSC was used to determine shrinkage, indicated by $T_s$, related to the thermal transitions of the collagens on heating and was employed in this study to investigate the effect telopeptide cleavage has on the thermal properties (Samouillan et al., 2011) (figure 2). Higher shrinkage temperatures (peak maximum) are typically indicative of a higher degree of intramolecular interactions between the collagen molecules. Telopeptide cleavage results in a smaller shrinkage temperature for both polymeric and monomeric collagen (table 1). This can be contributed to the covalent aldol cross-links provided by the C- and N- terminus to the adjacent molecule. Monomeric atelocollagen possesses a higher peak maximum compared to polymeric atelocollagen, despite the additional natural cross-links specific to polymeric collagen. This pattern is again portrayed with polymeric and monomeric collagen. The reason for this could be due to the loss of the triple helical conformation of polymeric collagen which was shown by the CD spectrum.

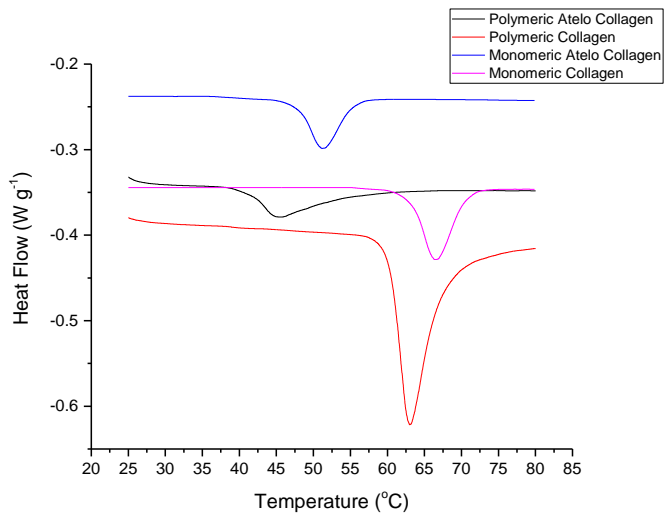

**Fig. 2.** DSC data for polymeric atelo, polymeric, monomeric atelo and monomeric type I collagen.

**Table 1.** DSC data from polymeric atelo, polymeric, monomeric atelo and monomeric type I collagen

| Collagen Type I | Average Enthalpy (W·g$^{-1}$) | Average Peak Maximum (°C) |
|---|---|---|
| Polymeric Atelo | -0.378 | 44.83 |
| Polymeric | -0.621 | 62.69 |
| Monomeric Atelo | -0.320 | 51.42 |
| Monomeric | -0.432 | 66.53 |

High pressure liquid chromatography (HPLC) was used to determine the amino acid content (residues per 1000 residues) in polymeric atelo and polymeric type I collagen (table 2). Hydroxyproline is formed intracellularly from the post-translational hydroxylation of proline and constitutes 10–14% of the total amino acid content of mature collagen (Seibel, 2005). The hydroxyproline content for both collagenous proteins was 9.66% of the total amino acid content (96.6 residues per 1000 residues) for polymeric collagen confirming to the literature value for mature (polymeric) collagen. Tyrosine is only present in the telopeptides of the collagen molecule and can be used as a measure of telopeptide removal {Menders, 2002 #203}. Existing normally as 0.5% of the total amino acid content, a total content below 0.2% (2 residues per 1000 residues) was found to serve as a suitable measure for sufficient removal of telopeptides (Herrmann and Mark, 1975, Cliche et al., 2003). The amino acid sequence of the telopeptides shown below for bovine type I collagen (Jones and Biophysics, 1987):

**The free α (1) N-terminal telopeptide conformation:**

α1    GLU-LE-SER-TYR-GLY-TYR-ASP-GLU-LYS-SER-THR-GLY-ILE-SER-VAL-PRO

**The free α (1) C-terminal telopeptide conformation:**

α1    SER-GLY-GLY-TYR-ASP-LEW-SER-PHE-LEU-PRO-GLN-PR-PPRO-GLN-GLX-LYS-ALA-HIS-ASP-GLY-GLY-ARG-TYR-TYR-ARG-ALA

**Table 2.** HPLC data from polymeric atelo and polymeric collagen displayed as residues per 1000 residues.

| Amino Acid | Atelo Polymeric Collagen (residues per 1000 residues) | Polymeric Collagen (residues per 1000 residues) |
|---|---|---|
| Aspartic Acid | 46.2 | 47.26 |
| Threonine | 16.2 | 16.36 |
| Serine | 34.1 | 33.19 |
| Glutamic Acid | 73.7 | 72.26 |
| Proline | 123 | 122.54 |
| Glycine | 337 | 334.00 |
| Alanine | 108 | 108.77 |
| Valine | 22.9 | 23.54 |
| Methionine | 6.31 | 6.61 |
| Isoleucine | 12.9 | 13.68 |
| Leucine | 25.23 | 26.26 |
| **Tyrosine** | **3.90** | **5.26** |
| Phenylalanine | 13.6 | 13.81 |
| Histidine | 5.70 | 5.85 |
| Lysine | 22.2 | 21.44 |
| Arginine | 52.0 | 52.57 |
| Hydroxyproline | **96.6** | **96.6** |

Polymeric collagen had a tyrosine total content of 0.53% and polymeric atelocollagen had a total content of 0.39%. This result is higher than the literature value of atelocollagen, however, the previous data from CD and DSC showed a definite difference between polymeric atelo and polymeric collagen. Instead, an explanation for the higher tyrosine occurrence could be as a result of telopeptide docking, whereby the free terminal structure docks onto the triple-helix chain as a staggered structure, so the tyrosine would still be accounted for in HPLC (Malone et al., 2004).

## 3.2 Physical characterization

SEM images were taken at varying magnifications to examine the internal material architecture (figure 3). Different morphologies are observed at 2000x and 8000x magnification after telopeptide cleavage.

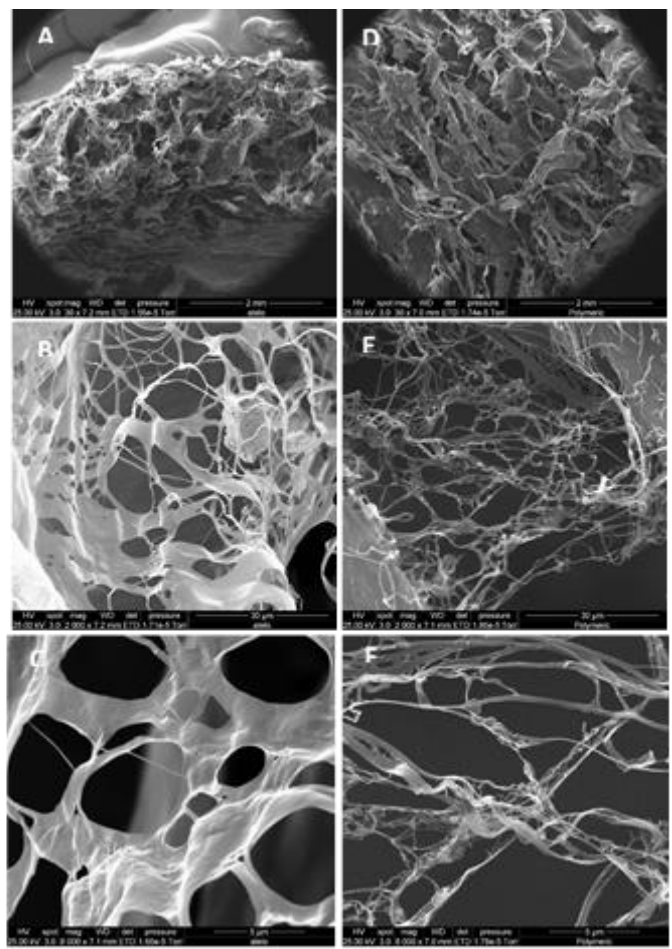

**Fig. 3.** SEM images: polymeric atelo collagen (A: 30 magnification, B: 2000 magnification, C: 8000 magnification) and polymeric collagen (D: 30 magnification, E: 2000 magnification, F: 8000 magnification).

## 4. Conclusions

The influence of telopeptides on the structure and physical properties of polymeric and monomeric was investigated. Monomeric atelocollagen displayed the characteristic positive band at 221 nm on the CD spectrum associated with the triple helical conformation of collagen. This implies that despite the covalent aldol cross-links provided by the telopeptides, their cleavage doesn't disrupt the natural collagen structure. It was found that polymeric collagen didn't display the characteristic positive peak at 221 nm, thereby implying that the natural cross-links associated with mature collagen disrupt the native triple helical structure. The influence of telopeptides was noted in the DSC results with decreased $T_s$ values after telopeptide cleavage for both monomeric and polymeric collagen, likely due to the reduced intramolecular aldol covalent cross-links attributed to the telopeptides. Telopeptide removal was confirmed using HPLC (decreased tyrosine content) with potential telopeptide docking.

## Acknowledgements

The author gratefully acknowledges the award of industrial research funding to R.A.H by DTC TERM. This work was supported by Southern Lights Biomaterials, New Zealand, Massey University, New Zealand and Collagen Solutions, UK.